\documentclass[aip,jcp,preprint]{revtex4-1}
\usepackage{graphicx}
\usepackage{rotating}
\usepackage{amsfonts,amsmath,latexsym,bbm,dsfont}
\usepackage{chemformula}
\usepackage[hidelinks]{hyperref}

\usepackage{color}

\newcommand{\NL}{\nonumber \\}

\newcommand{\beq}{\begin{eqnarray}}
\newcommand{\eeq}{\end{eqnarray}}

\newcommand{\op}{\hat}

\newcommand{\eq}[1]{Eq.~(\ref{#1})}

\newcommand{\fig}[1]{Figure~\ref{#1}}
\newcommand{\tab}[1]{Table~\ref{#1}}

\newcommand{\ASop}[2]{{\cal A}\left(#1;#2\right)}

\newcommand{\pyscf}{\textsc{PySCF} }
\newcommand{\elemcojl}{\textsf{ElemCo.jl} }
\newcommand{\quantwo}{\textsf{Quantwo} }
\newcommand{\Ms}{M_\mathrm{s}}
\newcommand{\avdz}{\textsc{avdz}}
\newcommand{\avtz}{\textsc{avtz}}
\def\RCS$#1: #2 ${\@namedef{RCS#1}{#2}\typeout{RCS #1: #2}}
\mathchardef\lt="313C \mathchardef\gt="313E
\mathcode`<="4268 \mathcode`>="5269

\renewcommand\vec\mathbf

\begin{document}

\author{Thomas Schraivogel}
\email{t.schraivogel@fkf.mpg.de}
\affiliation{Max Planck Institute for Solid State Research, Heisenbergstra\ss e 1, 70569 Stuttgart, Germany}
\author{Daniel Kats}
\email{d.kats@fkf.mpg.de}
\affiliation{Max Planck Institute for Solid State Research, Heisenbergstra\ss e 1, 70569 Stuttgart, Germany}

\title{Two Determinant Distinguishable Cluster}
\begin{abstract}
A two reference determinant version of the distinguishable cluster with singles and doubles (DCSD) has been developed.
We have implemented the two determinant distinguishable cluster (2D-DCSD) and the corresponding traditional 2D-CCSD method 
in a new open-source package written in Julia called \elemcojl.
The methods were benchmarked on singlet and triplet excited states of valence and
Rydberg character, as well as for singlet-triplet gaps of diradicals.
It is demonstrated that the distinguishable cluster approximation improves the accuracy of 2D-CCSD.
\end{abstract}

\maketitle 
\section{Introduction}
Coupled cluster (CC) theory\cite{shavitt09} provides us with a hierarchy of methods, which systematically converges to the exact solution of the
non-relativistic Schrödinger equation in the Born-Oppenheimer approximation in a given basis set.
Characteristic feature of the theory is the wave function ansatz based on the exponential of the excitation operator $\op T$ 
acting on a reference function $|0>$,
\begin{equation}
\Psi = e^{T} |0>.
\end{equation}
Due to the exponential wave function ansatz CC methods are naturally size-extensive, unlike configuration-interaction methods. 
The excitation operator in the cluster operator has to be truncated in practice to reduce the computational complexity.
Truncating it after the double excitation operator yields the coupled cluster method with single and double excitations (CCSD).
Single reference coupled cluster theory is using a single Slater determinant as reference function, on which the cluster operator acts upon.
However, in some wave functions there is no single determinant with a high enough weight to justify the single reference approach and 
a multireference coupled cluster (MR-CC) method is required. \newline
There are many different MR-CC methods with different strengths and weaknesses~\cite{lyakh12,koehn13,evangelista18}.
A useful high-level categorization is between Fock space and Hilbert space Ansätze\cite{lyakh12}.
Fock space methods operate in Fock space, a union of Hilbert spaces with different numbers of (quasi)particle each.
Hilbert space methods, on the other hand, operate in a single Hilbert space.
Hilbert space methods can further be subdivided into state-universal and state-specific\cite{koehn13} methods, depending whether they provide
several electronic states at once or one at a time.
We will later see that the two determinant coupled cluster method, the subject of this paper, is a state-universal Hilbert space method and arguably
the simplest of his kind.
The need for multireference coupled cluster methods stems from the truncation of the excitation operator.
In untruncated full CC, all determinants are explicitly parameterized in the CC wave function
and thus full CC corresponds to full configuration interaction (FCI) as long as the reference has a non-zero overlap 
with the desired wave function.
Multireference coupled cluster is significantly more difficult than single reference coupled cluster and because of that
many methods were developed in order to stay in the single reference framework as much as possible.
An incomplete list includes tailored coupled cluster\cite{kinoshita05}, the multireference coupled cluster method based on the single reference formalism 
by Oliphant and Adamowicz\cite{oliphant91} and the internally corrected coupled cluster methods\cite{paldus17}. \newline
Internally corrected coupled cluster methods emulate the effect of left-out excitation operators due to the truncation of the cluster operator
by appropriate modifications of the CC amplitude equations.
One of those methods is the distinguishable cluster (DC) approximation\cite{kats13}.
Distinguishable cluster theory\cite{kats13,kats19_dc,rishi19} has been shown to improve the accuracy and applicability 
of single reference coupled cluster considerably\cite{kats15,schraivogel21_dc}. \newline
Multiconfigurational wave functions are frequently encountered in the study of excited states.
One of the most common type of excited state is the open-shell singlet (OSS), 
for which two equally important reference determinants are required.
A method specifically developed for the calculation of OSS was born out of work on 
Hilbert space multireference coupled cluster theory\cite{kucharski91} in 1992 and 
was coined the two determinant coupled cluster with singles and doubles excitations (2D-CCSD) method\cite{balkova92}.
Excited states of ozone\cite{balkova92}, ketene\cite{balkova92}, diazomethane\cite{balkova92}, water\cite{balkova93}, ozone and butadiene\cite{szalay94}, cyclobutadiene\cite{balkova94}, 
methylene\cite{balkova95}, the dimers\cite{korkin96} of \ch{N2}, \ch{CO}, \ch{BF} and alkali metal dimers\cite{neogrady05} have been studied with 2D-CCSD.
Formulas for the energy gradient have been presented and used for geometry optimizations of ozone\cite{szalay94}, butadiene\cite{szalay94} and ketene\cite{szalay96}.
Non-iterative inclusion of perturbative triples resulting in 2D-CCSD(T) have been developed and applied in the cyclobutadiene\cite{balkova94} and methylene\cite{balkova95} study.
Unlike the canonical single reference CCSD method, 2D-CCSD is relatively sensitive to the choice of reference orbitals,
an effect intensively studied in 2018 on over 100 OSS excited states\cite{lutz18}.
Jeziorski and Paldus published in 1988 a linear two determinant coupled cluster method\cite{jeziorski88} and 
added quadratic terms shorty after\cite{paldus89}.
Albeit, instead of using two open-shell determinants, they used two closed-shell determinants and were
using an orthogonally spin adapted (OSA) formalism\cite{paldus77} instead of the spin orbital formulation used by Balkov\'a et. al.
Their OSA-2D-CCSD method has been combined with the ACP and ACPQ approximations\cite{piecuch93}, 
which are in turn related to the distinguishable cluster approximation. \newline
Traditionally, the most prominent route to excited states in coupled cluster theory is provided by the 
equation of motion\cite{emrich81,sekino84,geertsen89,stanton93} (EOM) or linear response\cite{monkhorst77,dalgaard83,koch90-1,koch90-2} framework.
In EOM-CC the excitation energies are obtained by a diagonalization of a non-Hermitian Jacobian, 
built from the Hamiltonian and a converged ground state CC wave function.
An EOM-DCSD theory has been developed and improved accuracy of excited states have been reported\cite{rishi17}.
The methods become problematic when already the ground state is not dominated by a single configuration, like in bond breaking or diradicals.
To overcome this shortcoming the spin flip approach was introduced\cite{krylov01} and combined with the CCSD method\cite{levchenko04} (EOM-SF-CCSD).
The high spin (single configurational) state is used as reference for the
initial CC calculation, while the multiconfigurational state is obtained by applying a spin flip operator on the resulting wave function,
which leads to a more balanced description of the states.
To avoid confusion, the canonical EOM-CCSD method will occasionally be denoted as EOM-EE-CCSD, with EE standing for excitation energy. \newline
Recently, alternatives to EOM within coupled cluster theory have been explored 
more \cite{mewes14,lee19,zheng19,zheng20,kossoski21,marie21,arias-martinez22,dreuw23,rishi23,tuckman23-1,tuckman23-2}.
In particular, the $\Delta$-CC approaches have received attention\cite{mewes14,lee19,zheng19,zheng20,kossoski21,arias-martinez22,dreuw23,rishi23}, 
especially in the study of core excited states\cite{lee19,zheng19,zheng20,arias-martinez22} and 
doubly (dark) excited states\cite{lee19,kossoski21,rishi23}, which are known to be problematic for EOM-CCSD.
We note in passing that a remedy for doubly excited states via the so called intermediate state approach\cite{ravi22} (IS-EOM) has been proposed.
In $\Delta$-CC approaches, two separate CC calculations are performed for the ground and excited state and the excitation energy is calculated as the
energy difference.
This offers the possibility to follow a state-specific\cite{burton22,marie23,kossoski23} philosophy and ameliorates biases towards the ground state.
The $\Delta$-CC approach takes advantage of the multiple solutions to the CC equations\cite{jankowski94-1,jankowski94-2,jankowski95,mayhall10}, 
which can be targeted by different starting guesses and reference functions.
The fact that one can converge a CC calculation on an excited state solution by a good enough starting point was already 
mentioned by Paldus et al. in the case of MR-CCSD, where the non-linear terms were found to be crucial for this endeavor\cite{paldus89}.
The single reference case was analyzed by Meissner et al.\cite{meissner93}.
Bartlett et al. calculated OSS with unrestricted single reference CC already in 1986, 
were they used an a-posterio correction to remove the spin-contamination stemming from the triplet\cite{magers86}. \newline
Recently, we have put forward the fixed-reference (FR) approach\cite{schraivogel21_dc} in order to calculate OSS with $\Delta$-CC.
In the fixed-reference approach the double amplitude which corresponds to the excitation of reference A to reference B 
of the OSS problem is fixed to one and all other amplitudes are allowed to relax in a normal CC calculation. 
We note that this simple idea was already mentioned in the literature \cite{shavitt09}, 
but to the best of our knowledge it was never thoroughly examined.
It was shown for a limited set of valence and Rydberg singlet and triplet excited states that the spin contamination is minimal and 
the accuracy of FR-CCSD is often comparable to EOM-CCSD.
Furthermore, FR-DCSD improves over FR-CCSD, and with the inclusion of triples excitations extremely accurate excited states 
can be obtained\cite{schraivogel21_dc}.
The implementation effort of the FR approach is negligible and it does not increase the cost of the computation.
While the distinguishable cluster approximation has proven to be extremely successful in single reference coupled cluster theory, its potential in
multireference coupled cluster theory is largely unexplored.
In this contribution we want to go the first steps along this line and present a new implementation of 2D-DCSD and some first benchmark calculations.

\section{Theory}
\begin{figure}
  \centering
  \includegraphics[scale=0.4]{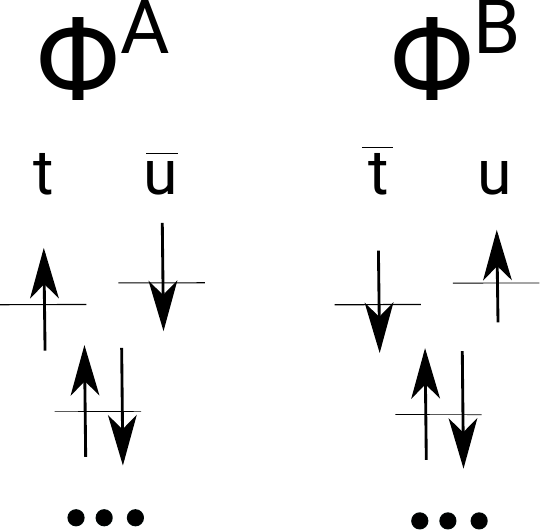}
  \caption{Two Slater determinants with two singly occupied orbitals each, which can be coupled to an open-shell singlet or $\Ms=0$ triplet CSF.
    The determinants are connected by spin flipping the electrons in the singly occupied orbitals. Note that the singly occupied orbitals
    are not degenerate in general. Degeneracies are frequently encountered in the study of diradicals. Overbar indicates beta spin.}
  \label{fig:oss}
\end{figure}

An open-shell singlet (or $\Ms = 0$ triplet) wave function has two leading determinants in its full configuration interaction (FCI) expansion with 
equal weights of same (opposite) sign.
The two determinants, let's call them $\Phi_A$ and $\Phi_B$, are displayed in \fig{fig:oss}
and are connected by flipping the spin of the two singly-occupied orbitals, called t and u, with overbar indicating beta spin.
Because of this symmetry the coupled cluster amplitudes normal-ordered with respect to determinant $\Phi_B$ ($\Phi_A$) can be expressed
in terms of amplitudes normal-ordered with respect to $\Phi_A$ ($\Phi_B$) by flipping the spin.
This relationship allowed Balkov\'a and Bartlett in 1992 to formulate and implement a two determinant coupled cluster theory \cite{balkova92} 
for the open-shell singlet problem in an effectively single reference way, 
building on previous work on Hilbert space multireference coupled cluster theory by Kucharski et al.\cite{kucharski91}.
Starting from the Bloch equation and using the Jeziorski-Monkhorst ansatz for the wave operator
the following energy and amplitude equations were derived\cite{balkova92,szalay94},
\begin{align}
\Delta E &= < {}^A\Phi| \op H_N e^{\op T_A} | {}^A \Phi > + < {}^A\Phi| \op H_N e^{\op T_B} | {}^B \Phi > \NL
         &\equiv \Delta E_A + W,
\end{align}
\begin{align}
R^i_a &= < {}^A\Phi^i_a |\op H_N e^{\op T_A} | {}^A \Phi >_C - \left( < {}^A\Phi^i_a | e^{\op T_B} | {}^B\Phi ><{}^B\Phi | \op H_N e^{\op T_A} | {}^A \Phi> \right)_C \NL
      &\equiv < {}^A\Phi^i_a | \op H_N e^{\op T_A} | {}^A \Phi >_C + M^i_a W = 0,
\end{align}
\begin{align}
R^{ij}_{ab} &= <{}^A\Phi^{ij}_{ab} |\op H_N e^{\op T_A}| {}^A \Phi>_C - \left( <{}^A \Phi^{ij}_{ab} |e^{\op T_B}| {}^B \Phi><{}^B\Phi |\op H_N e^{\op T_A}| {}^A\Phi> \right)_C \NL
            &-\ASop{ij}{ab} \Bigl[ <{}^A \Phi^i_a |e^{\op T_A}| {}^A\Phi> \left( <{}^A\Phi^j_b |e^{\op T_B}| {}^B\Phi> <{}^B\Phi |\op H_N e^{\op T_A}| {}^A\Phi>  \right)_C \NL
                           &- \op R(ia) <{}^B \Phi^i_a |e^{\op T_B}| {}^B\Phi> \left( <{}^A\Phi^j_b |e^{\op T_B}| {}^B\Phi> <{}^B\Phi |\op H_N e^{\op T_A}| {}^A\Phi> \right)_C \Bigr] \NL
            &\equiv <{}^A\Phi^{ij}_{ab} |\op H_N e^{\op T_A}| {}^A\Phi>_C + M^{ij}_{ab} W = 0,
\label{eq:2d-cc}
\end{align}
with the effective Hamiltonian $W = < {}^A\Phi| \op H_N e^{\op T_B} | {}^B \Phi > = <{}^B\Phi |\op H_N e^{\op T_A}| {}^A\Phi>$,
an antisymmetrization operator $\ASop{ij}{ab}$ antisymmetrizing the orbitals $i,j$ and $a,b$ in all possible ways 
and the operator $\op R(ia)$ to restrict the orbitals $i$ and $a$
to the doubly occupied and virtual space, in other words excluding the singly occupied orbitals.
The subscript $C$ indicates the connectedness of the terms.
The corresponding energy of the triplet can be obtained from the same calculation as $\Delta {}^{T\hspace{-0.08cm}}E = \Delta E_A - W$.
Focusing on the last line of the equations, the single reference coupled cluster equations plus renormalization terms $M_x W$ are recognized.
The only modification of the single reference CC equations is that we have to set the doubles amplitude, 
which corresponds to the excitation of ${}^A \Phi$ to ${}^B \Phi$ to zero.
The value of the corresponding element of the doubles residual is equal to $W$, and the element itself is also set to zero.
Working equations to build the $M$ tensor from the coupled cluster amplitudes can be found in Ref. \onlinecite{szalay94}.
Additionally, all implemented equations are documented in the repository of \elemcojl\cite{elemcojl}.
We note that the straightforward derivation based on \eq{eq:2d-cc} leads to terms containing the same spin-orbital twice, 
which we have omitted as in the previous implementations of 2D-CCSD, cf. Table~1 of Ref.~\onlinecite{szalay94}.
When the active orbitals $t$ and $u$ are of different spatial symmetry, then the active space consisting of $\Phi_A$ and $\Phi_B$ is complete (CAS).
The single excitations between the active orbitals are zero by symmetry, significantly reducing the number of terms.
Albeit, if the active orbitals fall in the same symmetry class, the active space is incomplete (IAS), 
because the two ionic configurations are missing and terms containing the all internal single amplitudes have to be included. \newline
The distinguishable cluster approximation was applied to the single reference doubles amplitude equations in the usual way.
The $M$ terms did not need to be changed with respect to the DC approximation.
From a user perspective, it does not matter whether $\Phi_A$ or $\Phi_B$ is chosen as the reference and we have verified that this is indeed the case
in our implementation.
\section{Computational Details}
\begin{figure}
  \centering
  \includegraphics[scale=1.0]{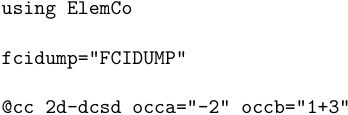}
  \caption{A minimal input script for \elemcojl to run a 2D-DCSD calculation for a four electron system using the FCIDUMP interface.}
  \label{fig:input}
\end{figure}

Electrons in core orbitals were not correlated.
Augmented basis sets of Dunning and coworkers of double (\avdz) and triple (\avtz) zeta quality have been used \cite{avnz}.
For second row atoms the respective refined basis sets (\textsc{av}(\textsc{n}+d)\textsc{z}) were utilized\cite{vn+dz}.
Basis sets and geometries are further specified in the respective sections and 
all used geometries can be found in the supplementary material.
If not stated otherwise, ROHF orbitals have been used for triplet states.
For open-shell singlet excited states MCSCF(2,2) natural orbitals optimized for the open-shell singlet CSF with restricted orbital
occupations\footnote{The orbital occupation restriction is only necessary if the active orbitals have the same spatial symmetry.} 
according to the reference CSF were employed,
which we will call in accordance with Ref.~\onlinecite{lutz18}, 2D-SCF orbitals.
The excited states have been calculated as energy differences between ground state and excited state CC energies.
Vertical excitations have been calculated throughout.
FR-DC-CCSDT calculations have been done with an automatically generated unrestricted DC-CCSDT implementation generated with our 
\quantwo\cite{quantwo} code generator and the
Integrated Tensor Framework \cite{shamasundar2011, MOLPRO-WIREs} in a development version of \textsc{Molpro} \cite{MOLPRO-WIREs, MOLPRO-JCP, MOLPRO}.
For the EOM-SF-CCSD calculations \pyscf\cite{pyscf1, pyscf2, libcint} was used. 
The semi-stochastic heat bath configuration interaction (SHCI) calculations were done with \textsc{Dice}\cite{dice1,dice2,dice3}.
FR-CC and 2D-CC calculations have been performed with \elemcojl \cite{elemcojl}.
All implemented equations can be found in the corresponding GitHub repository.
\elemcojl is an open-source Julia package actively developed in our group. 
It offers implementations of various electron correlation methods with a main focus on coupled cluster methods.
In this project we have used it with an FCIDUMP interface. 
The FCIDUMPs were generated after the ROHF or 2D-SCF calculation with \textsc{Molpro}\cite{MOLPRO-WIREs, MOLPRO-JCP, MOLPRO}.
An example \elemcojl script to perform a 2D-DCSD calculation for a four electron system is shown in \fig{fig:input}.
\section{Results}
As the first test we compare valence and Rydberg excited state energies with theoretical best estimates from the literature.
Furthermore, we calculate singlet-triplet gaps of several diradicals, including atoms, diatomics and the trimethylenemethane (TMM) molecule.
\subsection{Valence and Rydberg excited states}
\begin{figure}
  \centering
  \includegraphics[scale=1.0]{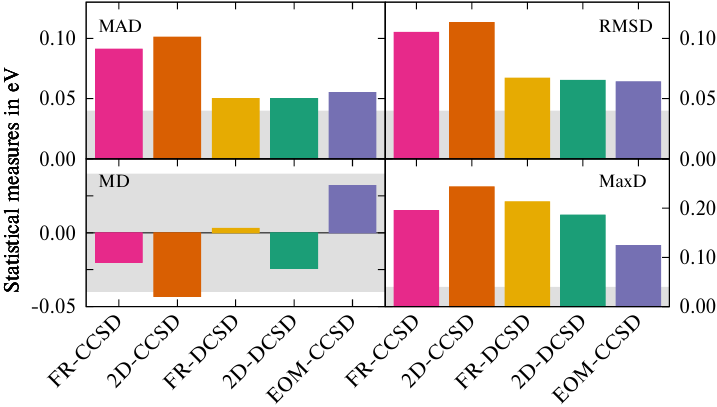}
  \caption{Statistical measures for the calculated (\avtz) excitation energies of \tab{tab:ee1} and \ref{tab:ee2}. 
           The region of chemical accuracy is shaded grey.
}
  \label{fig:stats}
\end{figure}

Singlet and triplet excited states of a handful of small molecules are presented in \tab{tab:ee1}.
The triplet excited states are easily accessible with single reference coupled cluster theory using the $\Ms=1$ component.
Nevertheless, we report the calculated $\Ms=0$ FR and 2D-CC values, which both differ from the $\Ms=1$ values as a sanity check.
They are compared to extrapolated selected CI calculations of FCI quality (exFCI) of Loos et al.\cite{loos18}.
Accounting for triple excitations can be crucial for some excited states and we included our results using
fixed-reference distinguishable cluster with
singles, doubles and triples (FR-DC-CCSDT) from Ref.~\onlinecite{schraivogel21_dc}, as well.
One excited state of butadiene and several excited states of ethylene are shown in \tab{tab:ee2},
where they are compared to EOM-CC values taken from the literature\cite{watson12, szalay17}. 
The butadiene calculation in aug-cc-pVTZ marks the biggest calculation performed with 22 electrons and 318 orbitals.
For this system the 2D-SCF optimization did not converge using the \avtz~basis\footnote{In \avdz~it converges without problems. In \avtz~it oscillates between a singlet and triplet solution and state-averaging both leads also to a convergent solution.} 
and because of that we used QRHF orbitals as described in Ref. \onlinecite{lutz18}. 
The statistics using the data of the two tables are visualized in \fig{fig:stats}.\newline
The excitation energies from the fixed-reference approach in combination with CCSD are quite accurate considering the simplicity of the approach,
with errors around 0.1 eV. The 2D-CCSD excitation energies are in general remarkably similar to the ones calculated with FR-CCSD. 
For the limited test set both methods showed almost the same performance.
The effect of the distinguishable cluster approximation is approximately the same in FR and 2D approaches 
and it halved the errors to around 0.05 eV, 
bringing them close to chemical accuracy of 0.04 eV and making FR-DCSD and 2D-DCSD on par with EOM-CCSD.
In some cases the FR formulation leads to erroneous results.
While the two determinant CC methods were able to resolve the dense excited states of ethylene nicely, the fixed-reference methods converged to
a different (unphysical) solution for the ${}^1B_{3u}$ state, which can not be fixed by taking triple excitations into account, which
is exemplified by the FR-DC-CCSDT value for the ${}^1B_{3u}$ state of 6.79 eV.
The same happened for the ${}^1B_{u}$ state of butadiene.
Using state-averaged orbitals, averaging the triplet and singlet, provides a possible remedy.
That leads for the problematic ${}^1B_{3u}$ state of ethylene to 7.87 and 7.91~eV and for the ${}^1B_{u}$ state of butadiene 
to 6.43 and 6.45~eV with FR-CCSD and FR-DCSD, respectively.
Those values have been used for the statistics displayed in \fig{fig:stats}.
The $\mathrm{^3\Sigma^+_u}$~excited state of the nitrogen molecule has four equally important zero order determinants in its wave function.
In our previous study it was seen that explicit triple excitations were needed to get somewhat satisfactory results with 
the fixed-reference approach~\cite{schraivogel21_dc}.
The two determinant methods show similar problems (albeit to a lesser degree) and this state was thus excluded from the statistics.
\begin{table}[htbp]
\caption{Vertical (\avtz) excitation energies in eV.
Geometries, EOM-CCSD and exFCI values from Ref.~\onlinecite{loos18}.
Fixed-reference values from Ref.~\onlinecite{schraivogel21_dc}.
The $\mathrm{^1A_1}$ state of ammonia and the $\mathrm{^3A_1}$ state of formaldehyde are IAS cases.}
\label{tab:ee1}
\begin{ruledtabular}
\begin{tabular}{cccccccc}
                            &  \multicolumn{3}{c}{CCSD} & \multicolumn{2}{c}{DCSD} & DC-CCSDT & exFCI      \\
                            & FR & 2D & EOM & FR & 2D & FR &  \\
\hline
\multicolumn{8}{c}{Nitrogen}      \\
$\mathrm{^1\Pi_g (n \rightarrow \pi^*)}$    & 9.37 & 9.42 & 9.41 & 9.28 & 9.31 & 9.33 & 9.34 \\
$\mathrm{^3\Sigma^+_u (\pi \rightarrow \pi^*)}$ & 8.15 & 8.24 & 7.66 & 8.69 & 8.08 & 7.57 & 7.70 \\
$\mathrm{^3\Pi_g (n \rightarrow \pi^*)}$    & 8.18 & 8.13 & 8.09 & 8.14 & 8.10 & 8.01 & 8.01 \\
\multicolumn{8}{c}{Carbon monoxide}                                          \\
$\mathrm{^1\Pi (n \rightarrow \pi^*)}$       & 8.61 & 8.62 & 8.59 & 8.58 & 8.56 & 8.48  & 8.49 \\
$\mathrm{^3\Pi (n \rightarrow \pi^*)}$       & 6.19 & 6.12 & 6.36 & 6.21 & 6.16 & 6.28 & 6.28 \\
\multicolumn{8}{c}{Water}                                                    \\
$\mathrm{^1B_1 (n \rightarrow 3s)}$     & 7.54 & 7.54 & 7.60 & 7.59 & 7.60 & 7.62 & 7.62 \\
$\mathrm{^1A_2 (n \rightarrow 3p)}$     & 9.31 & 9.32 & 9.36 & 9.37 & 9.37 & 9.40 & 9.41 \\
$\mathrm{^3B_1 (n \rightarrow 3s)}$     & 7.21 & 7.21 & 7.20 & 7.25 & 7.25 & 7.25 & 7.25 \\
$\mathrm{^3A_2 (n \rightarrow 3p)}$     & 9.19 & 9.18 & 9.20 & 9.24 & 9.24 & 9.24 & 9.24 \\
\multicolumn{8}{c}{Ammonia}                                                  \\
$\mathrm{^1A_1 (n \rightarrow 3s)}$     & 6.51 & 6.52 & 6.60 & 6.55 & 6.56 & 6.58 & 6.59 \\
$\mathrm{^1E (n \rightarrow 3p)}$        & 8.08 & 8.08 & 8.15 & 8.12 & 8.13 & 8.16  & 8.16 \\
\multicolumn{8}{c}{Formaldehyde}                                             \\
 $\mathrm{^1A_2 (n \rightarrow \pi^*)}$   & 3.82 & 3.83 & 4.01 & 3.90 & 3.91 & 3.96 & 3.98 \\
 $\mathrm{^1B_2 (n \rightarrow 3s)}$      & 7.18 & 7.19 & 7.23 & 7.20 & 7.21 & 7.22 & 7.23 \\
$\mathrm{^3A_2 (n \rightarrow \pi^*)}$    & 3.53 & 3.52 & 3.56 & 3.59 & 3.59 & 3.57 & 3.58 \\
$\mathrm{^3A_1 (\pi \rightarrow \pi^*)}$  & 6.05 & 5.97 & 5.97 & 6.09 & 5.96 & 6.06 & 6.06 \\
$\mathrm{^3B_2 (n \rightarrow 3s)}$       & 7.10 & 7.10 & 7.08 & 7.12 & 7.11 & 7.08 & 7.06
\end{tabular}
\end{ruledtabular}
\end{table}
\begin{table}[htbp]
\caption{Vertical singlet excitation energies in eV of butadiene and ethylene using the \avtz~basis.
Butadiene geometry and EOM reference values from Ref.~\onlinecite{watson12}. 
QRHF reference for butadiene with 2D and FR used.
Ethylene geometry and EOM reference values from Ref.~\onlinecite{szalay17}.
Using state-averaged orbitals solves the FR problems with the $\mathrm{^1B_u}$ and $\mathrm{^1B_{3u}}$ states (see also text).
The $\mathrm{^1B_{3u}}$ state of ethylene is an IAS case.
}
\label{tab:ee2}
\begin{ruledtabular}
\begin{tabular}{rcccccc}
                     & \multicolumn{3}{c}{CCSD} & \multicolumn{2}{c}{DCSD} & CCSDT \\
                     & FR & 2D & EOM & FR & 2D & EOM \\
\multicolumn{7}{c}{Butadiene} \\
$\mathrm{^1B_u}$     & 5.35   & 6.19 & 6.37      & 5.13   & 6.19   & 6.24     \\
\multicolumn{7}{c}{Ethylene} \\
$\mathrm{^1B_{1u}}$  & 7.22   & 7.23 & 7.41      & 7.30   & 7.31   & 7.35      \\
$\mathrm{^1B_{3u}}$  & 6.84   & 7.66 & 8.00      & 6.76   & 7.71   & 7.90      \\
$\mathrm{^1B_{3g}}$  & 7.90   & 7.90 & 8.06      & 7.98   & 7.98   & 8.02      \\
$\mathrm{^1B_{2g}}$  & 7.94   & 7.94 & 8.12      & 8.02   & 8.02   & 8.06      \\
$\mathrm{2^1B_{3g}}$ & 8.55   & 8.57 & 8.53      & 8.47   & 8.48   & 8.45      \\
$\mathrm{^1A_u}$     & 9.07   & 9.07 & 9.26      & 9.17   & 9.17   & 9.22      \\
$\mathrm{^1B_{1g}}$  & 9.87   & 9.88 & 9.77      & 9.73   & 9.74   & 9.70      \\
$2\mathrm{^1B_{2g}}$ & 9.94   & 9.96 & 10.01     & 9.97   & 9.96   & 9.93    
\end{tabular}
\end{ruledtabular}
\end{table}
\subsection{Spin splittings of diradicals}
Diradicals\cite{salem72,abe13,stuyver19} are a potential interesting application of the fixed-reference and two determinant methods. 
It is well known that some electronic structure methods struggle with certain diradicals 
because they overestimate the correlation of the triplet state in comparison to the singlet\cite{casanova20}.
Here, we benchmark the fixed-reference and two determinant methods on singlet-triplet gaps from the TS12 set 
assembled by Lee and Head-Gordon\cite{lee19two}.
We have added \ch{OH+} to the set and we will denote the resulting set TS13.
To test our methods on a bigger molecule we chose to calculate the $\mathrm{S}_1$-$\mathrm{T}_0$ gap of trimethylenemethane.
We found that using the $\Ms=1$ triplet energies leads to slightly better $\mathrm{S}_1$-$\mathrm{T}_0$ gaps, and therefore,
if not stated otherwise, the numbers correspond to this gap.
The difference in the studied systems was usually on the order of a few to at most a few tens of meV.
We compare to extrapolated semi-stochastic heat bath CI (ex-SHCI) and FR-DC-CCSDT calculations performed by us.
For perspective, we have performed and also present EOM-SF-CCSD calculations.
For the ex-SHCI calculations we started from the same orbitals as employed in our FR and 2D calculations and
for the singlet calculations we asked for two roots, providing us with the $\Ms=0$ triplet and OSS state. 
The results on the TS13 set are summarized in \tab{tab:stgaps}.\newline
The gaps for the atoms and diatomics are accurately reproduced by the fixed-reference and the two determinant methods, 
the vast majority lying within chemical accuracy.
The DC approximation again halves the error in both FR and 2D cases.
The FR-DCSD and 2D-DCSD methods perform slightly better than the EOM-SF-CCSD method.
Including triple excitations with the FR-DC-CCSDT method provides extremely accurate, almost FCI quality singlet-triplet gaps. \newline
The 2D-CC method, being a state universal Hilbert space method, has the attractive feature that the singlet and triplet energies can be obtained 
from one single calculation.
In general we opted for a state specific approach, because it was found that, "just as the optimal orbitals for the triplet states (ROHF) provide
the best TD-CCSD triplet energies, the optimal orbitals for OSS states (TD-SCF) provide the best TD-CCSD OSS energetics."\cite{lutz18}
\begin{table}[htpb]
\caption{Vertical $\mathrm{S_1-T_0}$ gaps in eV from the TS13 set. 
The \avdz~basis (\textsc{av}(\textsc{d}+d)\textsc{z} for second row atoms) have been used. 
Statistics relative to ex-SHCI in meV.
Bond lengths were taken from NIST Chemistry WebBook\cite{nist}.}
\label{tab:stgaps}
\begin{ruledtabular}
\begin{tabular}{lccccccc}
          & \multicolumn{3}{c}{CCSD}  & \multicolumn{2}{c}{DCSD} & DC-CCSDT & ex-SHCI \\
         & FR   & 2D   & EOM-SF & FR & 2D   & FR &  \\
NH       & 1.72 & 1.74 & 1.74 & 1.73 & 1.74 & 1.75 & 1.74 \\
\ch{OH+} & 2.33 & 2.35 & 2.35 & 2.34 & 2.35 & 2.36 & 2.36 \\
NF       & 1.63 & 1.66 & 1.63 & 1.62 & 1.64 & 1.57 & 1.58 \\
O2       & 1.10 & 1.12 & 1.11 & 1.06 & 1.07 & 1.08 & 1.07 \\
\ch{NO-} & 0.96 & 0.98 & 0.95 & 0.92 & 0.93 & 0.89 & 0.91 \\
C        & 1.41 & 1.43 & 1.42 & 1.42 & 1.43 & 1.45 & 1.44 \\
O        & 2.09 & 2.11 & 2.11 & 2.10 & 2.11 & 2.12 & 2.12 \\
PH       & 1.10 & 1.12 & 1.11 & 1.11 & 1.13 & 1.13 & 1.12 \\
PF       & 1.04 & 1.07 & 1.05 & 1.05 & 1.07 & 1.05 & 1.05 \\
S        & 1.26 & 1.29 & 1.28 & 1.28 & 1.30 & 1.30 & 1.30 \\
S2       & 0.62 & 0.66 & 0.65 & 0.60 & 0.63 & 0.63 & 0.63 \\
SO       & 0.85 & 0.88 & 0.85 & 0.83 & 0.84 & 0.83 & 0.82 \\
Si       & 0.91 & 0.93 & 0.92 & 0.93 & 0.94 & 0.96 & 0.96 \\
\hline
MD   & -6 & 20 & 5  & -9 & 7  & 0 \\
MAD  & 31 & 29 & 24 & 19 & 14 & 4 \\
RMSD & 34 & 40 & 28 & 22 & 21 & 7 \\
MAX  & 57 & 87 & 52 & 46 & 64 & 8
\end{tabular}
\end{ruledtabular}
\end{table}
As can be seen in \tab{tab:ncalcs} the singlet-triplet gaps calculated with 2D-CCSD and 2D-DCSD from a single calculation using
2D-SCF \emph{state-averaged} orbitals, mixing the OSS and corresponding triplet state, turned out to be a viable approach, reducing the number
of necessary calculations, without significant compromises in accuracy. \newline
\begin{table}[htpb]
\caption{Statistics for singlet-triplet gaps of the TS13 set calculated with 2D-CCSD and 2D-DCSD from a single calculation using
2D-SCF and ROHF state-averaged orbitals compared with values calculated from two different calculations using ROHF orbitals for the triplet and 
2D-SCF orbitals for the OSS. The \avdz~(\textsc{av}(\textsc{d}+d)\textsc{z}~for second row atoms) basis sets have been used. 
Statistics relative to ex-SHCI in meV.}
\label{tab:ncalcs}
\begin{ruledtabular}
\begin{tabular}{lcccc}
        & \multicolumn{2}{c}{2D-CCSD} & \multicolumn{2}{c}{2D-DCSD} \\
        & SA   & SS   & SA       & SS       \\
MD   & 17 & 20 & -2 & 7  \\
MAD  & 34 & 29 & 23 & 14  \\
RMSD & 46 & 40 & 32 & 21  \\
MAX  & 112 & 87 & 84 & 64 
\end{tabular}
\end{ruledtabular}
\end{table}
In \tab{tab:tmm} we compare calculated singlet-triplet gaps for a larger molecule, namely trimethylenemethane in a \avdz~basis (22e,142o) 
from FR and 2D CCSD and DCSD methods with FR-DC-CCSDT which was shown in this and 
previous studies \cite{schraivogel21_dc} to be an extremely accurate method.
TMM has in accordance to Hund's rule a triplet ground state and the first two singlet excited states of ionic and OSS character 
are degenerate in the ground state geometry. The degeneracy is lifted in the excited state geometry by symmetry distortion.
Going from left to right in \tab{tab:tmm} the difference between the $\Ms = 1$ triplet and the $\Ms = 0$ triplet calculated with FR and 2D methods
nicely decreases. 
While EOM-SF-CCSD overestimates the gap by 0.2 eV, all fixed-reference and two determinant methods perform better, with 2D-DCSD yielding
results very close to FR-DC-CCSDT with a discrepancy of 0.03 eV.
\begin{table}[htpb]
\caption{Singlet-triplet (\avdz) gap of the trimethylenemethane molecule. 
Additionally, the difference between the $\Ms = 0$ and $\Ms = 1$ triplet energies are shown as $\Delta\mathrm{T_0}$.
Ground state geometry of D3h symmetry of the triplet from Ref. \onlinecite{slipchenko02} used.}
\label{tab:tmm}
\begin{ruledtabular}
\begin{tabular}{ccccccc}
                      & \multicolumn{3}{c}{CCSD}  & \multicolumn{2}{c}{DCSD} & DC-CCSDT \\
                      & FR & 2D & EOM-SF  & FR & 2D & FR \\
$\mathrm{S_1-T_0}$ / eV            & 1.06   & 1.03  & 1.17        & 1.04   & 1.00   & 0.97       \\
$\Delta\mathrm{T_0}$ / meV & 128   & 65   &              & 47    & -11   & -0.1       
\end{tabular}
\end{ruledtabular}
\end{table}
\section{Conclusions}
The two determinant coupled cluster method has been combined with the distinguishable cluster approximation and was benchmarked
alongside the recently proposed fixed-reference approach.
For 25 excited states of seven small to medium sized closed-shell molecules the performance of 2D-CCSD and FR-CCSD was found to be remarkably similar,
both with errors around 0.1~eV.
The distinguishable cluster approximation halved the errors of both methods, bringing their performance to EOM-CCSD level, with errors around 0.05~eV.
For a diradical test set of thirteen atoms and diatomics the singlet-triplet gap was calculated with the FR and 2D-CC methods and compared to 
FCI quality benchmark data.
All methods yield $\mathrm{S}_1-\mathrm{T}_0$ gaps within chemical accuracy and the DC versions halved the errors of the corresponding CC methods.
It has been demonstrated that calculating the spin splittings in a single two determinant coupled cluster calculation with state-averaged orbitals
provides almost the same accuracy.
Singlet-triplet gaps from the FR-DC-CCSDT method are of almost FCI quality.
The singlet-triplet gap on a larger molecule, namely trimethylenemethane, was found to be accurately described by 2D-DCSD with an error of 0.03~eV,
while EOM-SF-CCSD was in error by 0.20~eV.

In conclusion, the fixed-reference and two determinant coupled cluster methods performance for spin splittings in
diradicals was seen to be quite remarkable, making them interesting approaches to study magnetic molecules \cite{kahn93,malrieu14}.
The two determinant coupled cluster methods and fixed-reference methods have been implemented in a new open-source Julia package called \elemcojl\cite{elemcojl}.
The package also provides implementations of other coupled cluster and mean-field methods and is actively developed in our group.
Among other features, non-Hermitian electron integrals can be used and we plan to use it in combination with transcorrelated 
methods\cite{boys69-2} in the future, potentially combining transcorrelation and the presented excited state methods.
\section*{Supplementary material}
See supplementary material for geometries and absolute energies.

\begin{acknowledgements}
Financial support from the Max-Planck Society is gratefully acknowledged.
\end{acknowledgements}

\section*{Author Declarations}
The authors have no conflicts to disclose.

\section*{Author contributions}
\textbf{Thomas Schraivogel:} Data curation (lead); Investigation (lead); Software (lead); Writing/Original Draft Preparation (lead); Writing/Review and Editing (equal). 
\textbf{Daniel Kats:} Conceptualization (lead); Writing/Review and Editing (equal); Software (supporting); Supervision (lead).

\section*{Data availability}
The data that support the findings of this study are available within the article and its supplementary material.

\bibliography{2d-dcsd,cc,cc_dc,cc_eom,cc_2d,cc_delta,cc_multiref,cc_tailored,books,magnetic,software,molpro,basis_sets,tc,papers}

\end{document}